\documentclass[twocolumn,showpacs,preprintnumbers,amsmath,amssymb]{revtex4}

\usepackage{graphicx}
\usepackage{dcolumn}
\usepackage{bm}

\begin{document}


\title{Design of coupling for synchronization in time-delayed systems}

\author{Dibakar Ghosh$^1$}
\author{Ioan Grosu$^{2,3}$}%
\author{Syamal K. Dana$^4$}

\affiliation{%
$^1$ Department of Mathematics, University of Kalyani, West Bengal 741235, India\\
$^2$ Faculty of Bioengineering, University of Medicine and Pharmacy, {``}Gr. T. Popa," Iasi, Romania\\
$^3$ Faculty of Chemistry, AI.I.Cuza University, 700506 Iasi, Romania\\
$^4$ Central Instrumentation, CSIR-Indian Institute of Chemical Biology, Jadavpur, Kolkata 700032, India.
}%

\date{\today}

\begin{abstract}
We report a design of delay coupling for targeting desired synchronization in delay dynamical systems. We target synchronization, antisynchronization, lag-, antilag- synchronization, amplitude death (or oscillation death) and generalized synchronization in mismatched oscillators. A scaling of the size of an attractor is made possible in different synchronization regimes. We realize a type of mixed synchronization where synchronization, antisynchronization coexist in different pairs of state variables of the coupled system. We establish the stability condition of synchronization using the Krasovskii-Lyapunov function theory and the Hurwitz matrix criterion. We present numerical examples using the Mackey-Glass system and a delay R\"{o}ssler system.
\end{abstract}


\maketitle
{\bf Many dynamical systems in physics, biology and engineering are often modeled by delay differential equations. In lasers, a light signal takes a finite time to travel an optical path to interact with similar devices. In the brain,  populations of neurons at distant locations interact through exchange of signals which take a finite time to travel. In electronic circuits, inductors and transmission lines delay a signal during transmission to another end. Studies of synchronization in delay systems consider an extrinsic coupling delay besides the intrinsic delay in the model system. In conventional studies of synchronization [1-2], the coupling function is usually assumed  {\it a priori} known. A region of system parameters and a coupling strength of the coupled system is then searched for where the stability of synchronization is ensured. Alternatively, engineering of synchronization is attempted [3-8], in recent times, that assumes the coupling function as {\it a priori} unknown. It decides a target synchronization state in a set of given dynamical systems and then formulate the coupling function to realize the desired coherent state whose stability is ensured by a suitable stability measure of the error dynamics. The benefit of engineering of synchronization is manifold: one can target many desirable coherent states in coupled systems which are usually not realizable by the conventional methods and it allows a control of synchronization. So far these methods of coupling design mostly deal with non-delay systems. We extend this engineering approach  to delay coupling in delay dynamical systems. }

\section{Introduction}
\par Time delay is an inherent property of many dynamical systems in nature [9-10] and in practical systems such as optoelectronic devices and electronic systems. As example, consider a time-delayed system i.e., $\dot{x}=f(x(t), x(t-\tau))$ where $\tau$ is a delay time. This intrinsic time delay makes a system infinite dimensional and more complex. In addition, a coupling delay may appear during a transmission of signal or information in natural systems like long axons of neurons [11], practical systems like communication channel via optical-link [12], electrical transmission lines. This transmission delay induces additional instabilities in the process of an emergent dynamics in coupled delay systems. As a result, studies of synchronization in delayed systems become more involved compared to those in non-delay systems [1-2] and they were therefore given a special attention [9-10, 13]. Different forms of coupling, namely, delay diffusive coupling [10, 13, 14], synaptic delay coupling [15] or special type of delay coupling [16-17] were used. The coupling configuration was always assumed known {\it a priori}; the coupling strength was played with to achieve a stable synchronization state. Stability of synchronization in delayed systems under delay coupling was mainly established using the Krasovskii-Lyapunov function theory [18] that showed dependence on system parameters and a coupling strength. However, these methods cannot realize all desirable coherent states in any dynamical system such as antisynchronization that is observable in inversion-symmetric systems only [19, 20].
\par It is important, for practical purposes, to search for a unified delay coupling strategy that can realize all desirable synchronization regimes in delay systems without restriction. It shall ensure stability of synchronization under parameter mismatch and if possible, the stability criterion is to be independent of the system parameters which is necessary to enable a control of synchronization. We attempt to address the issues, in this paper, and propose a general mathematical definition of unidirectional delay coupling for delay dynamical systems to target a desired synchronization state and to control it. The coupling configuration is assumed  {\it a priori} unknown, and its exact form is only derived from the definition of a dynamical system based on a general stability measure of the error dynamics and applicable for a large class of systems. Engineering of synchronization is explored [3-8] recently in  nonlinear dynamical systems to target many desired coherent states/patterns but they mostly dealt with non-delay systems. On the other hand, the authors reported an open-plus-closed-loop (OPCL) scheme [20-23] of coupling design for chaotic systems to realize a targeted state of synchrony by ensuring the stability of synchronization based on the Hurwitz matrix criterion [24]. An expression of coupling strength did not appear explicitly in the coupling definition. The coupling function is first designed from the given model system and then it is inserted between the dynamical systems to observe a desired synchronization state such as complete synchronization [20], antisynchronization [21], lag synchronization [22], generalized synchronization [23] and even a state of amplitude death (AD) [25] or oscillation death (OD) [26]. It is worth mentioning that the conventional diffusive coupling have some restrictions: antiphase is observed only in inversion symmetric systems [19, 20], LS in mismatched systems [27], GS in nonidentical or largely mismatched systems [28]. The OPCL coupling overrules [20] these restrictions, however, this method is so far restricted to non-delay systems.
\par We utilize the general benefits of the OPCL coupling scheme and extend it to delay dynamical systems in presence of coupling delay. The OPCL method is used recently [29] for synchronization in time-delay systems based on the Hurwitz matrix stability. We also derive the stability condition of synchronization in time delay system using the Hurwitz matrix stability criterion [24] but additionally support it by the Krasovskii-Lyapunov function theory [18].  The stability is now independent of the system parameters and valid for mismatched oscillators too. The stability criterion being free from the system parameters, a control of synchronization becomes viable now. By control of synchronization, we mean a strategy to change one to the other form of synchronization continuously by varying a control parameter without loss of stability during a transition. We have physically implemented [30] this control of synchronization in electronic circuit by inserting a scaling parameter in the definition of instantaneous coupling in nondelay system. Here we show that this control strategy can be applied to delayed systems under delay coupling. This is in contrast to the conventional methods [9-10, 13, 15-17] of delay coupling where such a flexible control of synchronization of different regimes is not possible in time delay systems. As a result, once the delay coupling is designed and switched ON, we are able to achieve several synchronization scenarios, CS, AS, LS, anti-lag synchronization (anti-LS) by controlling the scaling parameter. Using this external parameter, a scaling of the size of a driver attractor (amplification or attenuation) is possible at a response system. The GS is realized using a delay coupling with a matrix type scaling function, which is a general form of the scaling parameter. Additionally, we are able to engineer a mixed type of synchronization (MS) [20] in the response oscillator when similar pairs of state variables of the coupled systems attain separate synchronization states, CS and AS. Even an amplitude death (AD) [25] can be induced in the response system by stabilizing its equilibrium at origin. We support the theory with numerical simulations of Mackey-Glass system [31] and a time-delayed R\"{o}ssler system [32]. This coupling is not difficult to implement in practical systems, opto-electronic and electronic experiments and hence opens up doors for practical applications.
\par The rest of the paper is organized as follows: a general theory of the OPCL delay coupling in delay systems is discussed in Section II. The stability condition of synchronization is derived for mismatched systems. In Section III, numerical examples of the Mackey-Glass system and the time-delayed R\"{o}ssler system are described for instantaneous and delay coupling. The GS is elaborated in section IV. Results are concluded in Section V.
\section{Theory of OPCL delay coupling}
\par We consider a drive-response type unidirectional coupling. The driver is,
$$\dot y=f(y, y_\tau, \eta)+\Delta f(y, y_\tau, \eta),\;\;\;\;\; y, y_\tau \in R^n \eqno{(1)}$$
where $y_\tau =y(t-\tau ), $ $\eta$ is the vector of parameters and $\Delta f(y, y_\tau, \eta)$ denotes the parameter mismatch. If all parameters appear in the linear terms in $f(\cdot )$ then $\Delta f(y, y_\tau, \eta)=f(y, y_\tau, \delta \eta)$, otherwise $\Delta f(y, y_\tau, \eta)=f(y, y_\tau, \eta+\delta \eta)-f(y, y_\tau, \eta)$, in general, when $\delta \eta$ denotes the parameter mismatch. It drives a response system $\dot x=f(x, x_\tau, \eta),\; x, x_\tau  \in R^n,$ to induce a desired goal dynamics $g(t)=\alpha y(t-\tau_c);$ $\alpha=(\alpha_{ij})_{n\times n}$ is a constant matrix called as the scaling matrix and $\tau_c\geq 0$ is the coupling delay. The response system after coupling is,
$$\dot x=f(x, x_\tau, \eta)+D (x, x_\tau, y, y_\tau) \eqno{(2)} $$
where $D$ is a coupling term to be defined or designed. The error signal of the coupled system is defined by $e=x-\alpha y_{\tau_c}$. The $f(x, x_{\tau}, \eta)$ can be written as
$$f(x, x_{\tau}, \eta)=f(\alpha y_{\tau_c}+e, \alpha y_{\tau+\tau_c}+e_{\tau}, \eta)=f(\alpha y_{\tau_c}, \alpha y_{\tau+\tau_c}, \eta)$$$$
+\frac{\partial f}{\partial(\alpha y_{\tau_c})}e+\frac{\partial f}{\partial(\alpha y_{\tau+\tau_c})}e_\tau + \cdot \cdot \cdot  \eqno{(3)}$$
Keeping the first order terms in (3), the error dynamics is approximated,
       $$\dot e=\dot x-\alpha  \dot y_{\tau_c} \hskip 200pt$$
$$=f(\alpha y_{\tau_c}, \alpha y_{\tau+\tau_c}, \eta)+\frac{\partial f}{\partial(\alpha y_{\tau_c})}e+\frac{\partial f}{\partial(\alpha y_{\tau+\tau_c})}e_\tau+D-\alpha \dot y_{\tau_c}$$
$$=H_1e+H_2 e_\tau \hskip 200pt \eqno{(4)}$$
where
$$D=\alpha \dot y_{\tau_c}-f(\alpha y_{\tau_c}, \alpha y_{\tau+\tau_c}, \eta) +(H_1-\frac{\partial f}{\partial(\alpha y_{\tau_c})})e$$$$+(H_2-\frac{\partial f}{\partial(\alpha y_{\tau+\tau_c})})e_{\tau} \eqno{(5)}$$
is defined as the OPCL delay coupling and $H_1, H_2$ are two $n \times n$ arbitrary constant matrices.
\par To establish stability of synchronization, we first apply the Krasovskii-Lyapunov function [18],
$$V(t)=\frac{1}{2}e^T e+ \gamma  \int_{-\tau}^0 e^T(\theta ) e(\theta ) d\theta  \eqno{(6)}$$
where $\gamma$ is a positive constant.
Taking time derivative of (6) and using the error equation (4),
$$\dot V=\frac{1}{2}[(e^T H_1^T+e_\tau^T H_2^T)e+e^T(H_1 e+H_2 e_\tau)]+{\gamma}  [(e^T e-e_\tau^T e_\tau)]$$
$$= \left( \begin{array}{c}
  e \\
  e_\tau
\end{array} \right)^T \left( \begin{array}{cc}
  \frac{1}{2}(H_1^T+H_1)+\gamma  I_{n\times n} & \frac{1}{2}H_2 \\
  \frac{1}{2}H_2^T  & -\gamma  I_{n\times n}
\end{array} \right)  \left( \begin{array}{c}
  e \\
  e_\tau
\end{array} \right) $$
$$=E^T P E \hskip 200pt  \eqno{(7)}$$
where
$$E=\left( \begin{array}{c}
  e \\
  e_\tau
\end{array} \right),\;\; P= \left( \begin{array}{cc}
  \frac{1}{2}(H_1^T+H_1)+\gamma  I_{n\times n} & \frac{1}{2}H_2 \\
  \frac{1}{2}H_2^T  & -\gamma  I_{n\times n}
\end{array} \right) $$
We now prove that $\dot V<0$ if $P<0$. Let $Q$ and $R$ be two symmetric matrices and a matrix $S$ has a suitable dimension, then
$$\left( \begin{array}{cc}
  Q & S \\
  S^T  & R
\end{array} \right)<0$$
if and only if both $R<0$ and $Q-S R^{-1} S^T<0.$
\par From the above results we find that $\dot V<0$ if
$$H=2(H_1+H_1^T)+4\gamma  I_{n\times n}+\frac {1} {\gamma}  H_2 H_2^T<0  \eqno{(8)}$$
The asymptotic stability of synchronization is now ensured provided the elements of the $H_1$ and $H_2$ matrices are so appropriately chosen that the condition (8) is satisfied. We construct $H_1$ and $H_2$  from the {\it Jacobian} matrices of the system, $\frac{\partial f}{\partial(\alpha y_{\tau_c})}$ and $\frac{\partial f}{\partial(\alpha y_{\tau+\tau_c})}$. If the elements of the {\it Jacobian} matrices contain state variables, we replace them by constants $h_{i}$ keeping other elements (constants or zeros) unchanged. On the other hand if there is no state variable in the {\it Jacobian}, we replace some of the zero elements by constants $p_{i}$, keeping other constants as it is. For a $3\times 3$ {\it Jacobian}, the characteristic equation of a $H$ matrix is $\lambda^3+a_1 \lambda^2+a_2 \lambda+a_3=0 $ where its coefficients $a_i (i=1, 2, 3)$ are defined by the elements of the matrix. The corresponding Routh-Hurwitz (RH) criterion [24] is given by $a_1>0, a_3>0, a_1 a_2>a_3$ when the $H$ matrix has eigenvalues with all negative real parts. In fact, the parameter values $h_{i}$ and $p_{i}$ in $H_1$ and $H_2$ are so selected that the RH criterion is satisfied. The condition (8) is thereby fulfilled and ensures stability of synchronization. To complete the design of the coupling $D$, the $\alpha$ matrix and the coupling delay $\tau_c$ remains to be decided. The stability condition (8) clearly depends upon the choices of $h_{i}$ and $p_{i}$ but independent of the system parameters and the choice of $\alpha$ matrix and $\tau_c.$ This incorporates a great flexibility in this coupling design in realization of several synchronization scenarios, CS, AS, LS, anti-LS, MS and GS in identical or mismatched systems and a control from one to the other form of synchronization without intermediate loss of synchronization.

\par As in many practical delay dynamical systems, $x, y\in R^1$, when $\alpha $ is a scalar parameter, CS and AS can be induced simply by taking  $\alpha =1, -1$ respectively. Since $\alpha$ is not involved in the stability condition of synchronization, it can be used for a smooth control of synchronization from CS to AS or vice versa and even for inducing a flip-flop between CS ($\alpha =1$) and AS ($\alpha =-1$) states, which can be of practical use in digital encoding. A scaling of the size (amplification or attenuation) of the driver attractor at the response oscillator is possible by taking $\mid \alpha \mid >1 $ or $\mid \alpha \mid <1$. In higher dimensional delay systems ($x, y\in R^n$) too, we can realize all the known synchronization scenarios and in addition, we can target GS and MS when $\alpha$ is considered as a matrix, instead of a scalar parameter. If $\alpha$ is a diagonal matrix with different constant values (positive, negative or zero), a MS state is realized. In addition, if the off-diagonal elements of the $\alpha$-matrix has nonzero values, a GS state is obtained. We demonstrate all these synchronization events using numerical examples in the next section.

\section{Targeting synchronization: Numerical Examples}
\par  We explain the design of the delay coupling using first mismatched Mackey-Glass systems,

\par $$\dot y=-(\eta_1  + \delta \eta_1)y+\frac{(\eta_2+ \delta \eta_2)y_{\tau}}{1+y_{\tau}^{10}} \eqno{(9a)}$$
  $$\dot x=-\eta_1 x+\frac{\eta_2 x_{\tau}}{1+x_{\tau}^{10}}+D \eqno{(9b)}$$
which are chaotic for a set of parameters $\eta_1=1, \eta_2=2, \tau=2$ and  mismatches $\delta \eta_1=0.2, \delta \eta_2=0.3.$  The response system (9b) after adding the delay coupling (5) is
$$\dot x= - \delta \eta_1 \alpha y_{\tau_c}+\frac{\alpha  (\eta_2+\delta \eta_2)y_{\tau+\tau_c}}{1+y_{\tau+\tau_c}^{10}}-\frac{\eta_2 \alpha y_{\tau+\tau_c}}{1+\alpha ^{10} y_{\tau+\tau_c}^{10}}$$$$+(H_1+\eta_1)(x-\alpha y_{\tau_c})+[H_2-\eta_2 \frac{1-9\alpha ^{10} y_{\tau+\tau_c}^{10}}{(1+\alpha ^{10} y_{\tau+\tau_c}^{10})^2}](x_{\tau}-\alpha y_{\tau+\tau_c}) \eqno{(10)}$$
For this example, $H_1$ and $H_2$ are scalars, the condition for stability is derived from (8) when $\gamma =1$,
$$H_1+1+\frac{H_2^2}{4}<0 \eqno{(11)}$$
\begin{figure}
\includegraphics[width=90mm,height=65mm]{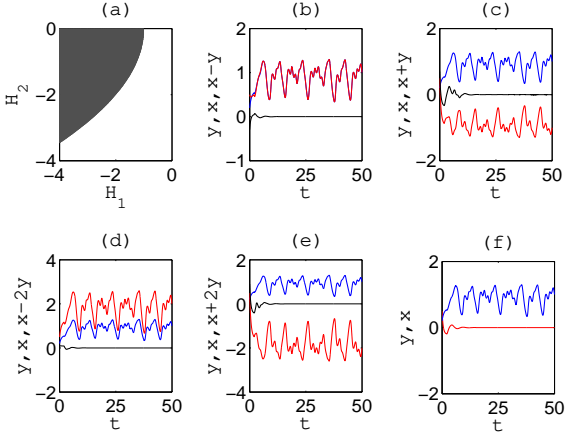}
\caption{\label . (Color online) Synchronization region in the $H_1-H_2$ plane in black in (a) for Mackey-Glass system. For $\tau_c=0,$ a pair of time series of driver (online blue) and response (online red) and, the synchronization error in black lines are shown, (b) CS for $\alpha =1$, (c) AS for $\alpha =-1,$ (d) CS with amplification for $\alpha =2,$ and (e) AS with amplification for $\alpha =-2$, (f) response at amplitude death for $\alpha =0.$ }
\end{figure}

The stability region for synchronization in the $H_1-H_2$ plane is numerically determined as shown in the black region in Fig.1(a). We make a choice of $H_1=-2, H_2=-1$ for our simulations. We first discuss the case with zero coupling delay, $\tau_c=0$. Figure 1(b) shows that the pair of time series of the driver $y(t)$ and the response $x(t)$ are in CS for $\alpha =1$ since the synchronization error e(t)=$x(t)-y(t)$ in black line converges to zero after the transients. For $\alpha =-1,$ the AS state between $x(t)$ and $y(t)$ is clear in Fig. 1(c) when the synchronization error e(t)=$x(t)+y(t)$ converges to zero in black line. Figure 1(d) shows a CS state with amplification ($\alpha>1$) of the driver at the response while in Fig.1(e), an AS state is seen with amplification $(|\alpha|>1)$ of the driver again at the response. Their corresponding errors at zero are shown in black lines. Similarly, attenuation can be observed by simply taking $0<\mid \alpha \mid<1$ but we do not elaborate here. It is also interesting that AD can be induced in the response system by taking $\alpha =0$ as shown in Fig.1(f). The driver $y(t)$ is oscillatory while the response $x(t)$ ceases to oscillate and is stable at zero. The  coupling actually stabilizes the equilibrium origin of the response system. In fact, the coupling has an inherent property to create a new equilibrium at origin if it is not present in the uncoupled system and then to stabilize it for a choice of $(\alpha=0)$. This is one of the major advantages of this coupling over the conventional delay coupling and can be utilized for the purpose of stabilizing a practical delay system by an external driving system.

\begin{figure}
\includegraphics[width=90mm,height=70mm]{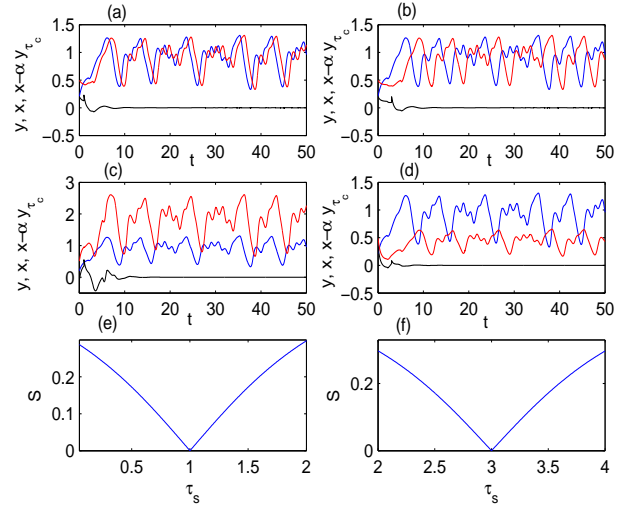}
\caption{\label. (Color online) Lag synchronization in Mackey-Glass system: time series of driver (blue) and response (red) for (a) $\alpha =1, \tau_c=1$, (b) $\alpha =1.0, \tau_c=3.0$. LS with (c) amplification for $\alpha =2, \tau_c=1$ and (d) attenuation for $\alpha =0.5, \tau_c=3.$  Error dynamics (black line) converges to zero after initial transients for all the cases. Similarity measure $S$ has a global minimum at (e) $\tau_{s0}=1$, (f) $\tau_{s0}=3$. }
\end{figure}
\par For nonzero coupling delays $\tau_c$, systems (9) and (10) develop a LS state instead of a CS state. In a LS state, all the state variables of the oscillators maintain amplitude correlation but shifted by the desired coupling delay $\tau_c$. Figure 2 plots the pair of time series of the response $x(t)$ (online red color) and the driver $ y(t)$ (online blue color) in LS for two different choices of time lag $\tau_c=1.0, 3.0.$ Figures 2(a) and 2(b) in upper panels show LS with $\alpha =1$ and  (a) $\tau_c=1.0,$  (b) $\tau_c=3.0$ respectively. The LS scenarios is clear from Figs. 2(a) and 2(b) by visual check. On the other hand, Figures 2(c) and 2(d) show LS with amplification ($\alpha =2$) and attenuation ($\alpha =0.5$) with coupling delay $\tau_c=1.0$ and $\tau_c=3.0$ respectively. In this case, LS scenarios are not clear from a visual check. Hence to confirm LS in the pair of time series, $y(t)$ and $x(t)$, we use a similarity measure (S)[1, 27],
$$S=\frac{<[x(t)-\alpha y(t-\tau_s)]^2>}{\sqrt{<x^2(t)> <\alpha^2 y^2(t)>}}.$$
The $S$ plot with arbitrary $\tau_s$ shows a global minimum at zero for a $\tau_s=\tau_{s0}$ which is the principal lag characteristic of a pair of time series[27].  Figure 2(e) shows a plot $S$ for the pair of $y(t)$ and $x(t)$ in Fig.2(c), which has a global minimum at $\tau_c=1.0=\tau_{s0}$ and confirms the LS of $\tau_c=1.0$. A similar LS scenario of larger time lag $\tau_c=3.0=\tau_{s0}$ is shown in Fig.2(f) for the pair of time series in Fig.2(d). Thus we are able to induce arbitrarily desired lag $\tau_c$ in the drive-response system, one smaller and another larger and, in addition, we amplify/attenuate the drive signal at the response, which is another major advantage over the conventional coupling. We can also stabilize the response system at equilibrium origin by simply taking $\alpha =0$ but we do not elaborate again. Note that we choose $\tau_c$ values one larger and another lower than the intrinsic time lag $\tau$ and, LS is not affected by this choice. The efficacy of a coupling design for larger coupling delay is, particularly, important in regards to synchronization in distant regions of the brain with large conducion delays [33].

\par As a second example, we consider a time-delayed R\"{o}ssler system [32]. The driving system is
$$\dot y_1=-y_2-e^{y_3(t-\tau )} \eqno{(12a)}$$
$$\dot y_2=y_1+a y_2 \eqno{(12b)}$$
$$\dot y_3=y_1+(b+\delta b)e^{-y_3(t-\tau )}-c \eqno{(12c)}$$
where $\delta b$ is a mismatch. The driver system is chaotic for $a=0.4, b=2.0, c=4.0$ and $0\leq \tau \leq 0.35$. In our simulation we consider $\tau=0.1$ and a parameter mismatch $\delta b=0.05.$

After adding the OPCL coupling (5), the response system becomes,
$$\dot x_1=-x_2-e^{x_3(t-\tau )}-\alpha_{11} \{y_2(t-\tau_c)+e^{y_3(t-\tau-\tau_c )}\}
     $$$$+\alpha _{22} y_2(t-\tau_c)+e^{\alpha _{33} y_3(t-\tau-\tau_c )}+p_1\{x_1-\alpha_{11} y_1(t-\tau_c)\}$$$$
+\{h_1+e^{\alpha_{33} y_3(t-\tau-\tau_c )}\}\{x_3(t-\tau)-\alpha_{33} y_3(t-\tau-\tau_c)\}\eqno{(13a)}$$
 $$\dot x_2=x_1+ax_2+(\alpha_{22}-\alpha_{11}) y_1(t-\tau_c)+p_2\{x_2-\alpha_{22} y_2(t-\tau_c)\}\hskip 200pt \eqno{(13b)}$$
$$\dot x_3=x_1+b e^{-x_3(t-\tau )}-\alpha _{33} c+(\alpha _{33}-\alpha_{11}) y_1(t-\tau_c) \hskip 100pt$$$$+\alpha_{33}(b+\delta b)e^{-y_3(t-\tau-\tau_c )}-b e^{-\alpha_{33} y_3(t-\tau-\tau_c )}$$$$+\{h_2+be^{-\alpha_{33} y_3(t-\tau-\tau_c )}\}\{x_3(t-\tau)-\alpha_{33} y_3(t-\tau-\tau_c)\}$$$$+p_3\{x_3-\alpha_{33} y_3(t-\tau_c)\} \eqno{(13c)}$$
where we choose $\alpha=[\alpha_{11} \;\;0 \;\;0\;;\; 0\;\; \alpha_{22}\;\; 0\;;\; 0\;\; 0\;\; \alpha_{33}]^T$ as a diagonal matrix.
In this case, $x, y\in R^3$, and hence $H_1$ and $H_2$ matrices are $3 \times 3$ matrices. The {\it Jacobian} matrices of the system (12) are
$$ \frac{\partial f}{\partial y} = \left( \begin{array}{ccc}
  0 & -1 & 0 \\
  1 & a & 0 \\
  1 & 0 & 0
\end{array} \right)\; \mbox{;}\; \frac{\partial f}{\partial y_{\tau}} = \left( \begin{array}{ccc}
  0 & 0 & -e^{y_3(t-\tau)} \\
  0 & 0 & 0 \\
  0 & 0 & -b e^{-y_3(t-\tau)}
\end{array} \right)$$
\begin{figure}
\includegraphics[width=50mm,height=40mm]{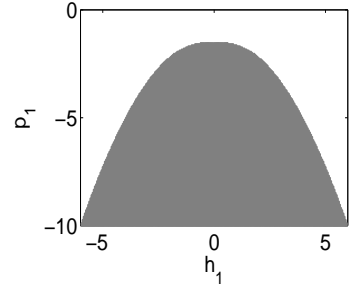}
\caption{\label. Synchronization region in the black region in $h_1 - p_1$ plane for delay R\"{o}ssler system. $p_1 = p_2=p_3$}
\end{figure}
\begin{figure}
\includegraphics[width=90mm,height=120mm]{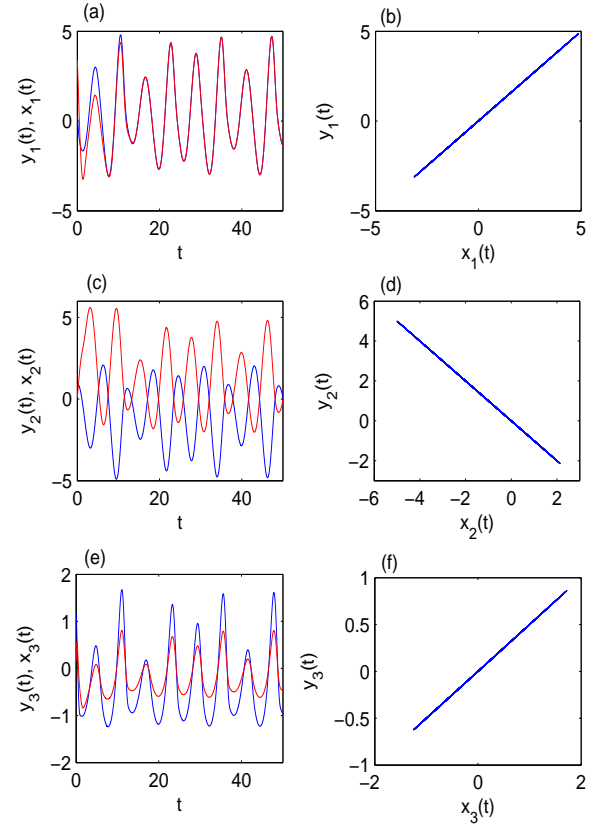}
\caption{\label. (Color online) Mixed synchronization in delay R\"{o}ssler system: CS between $y_1$ and $x_1$ for $\alpha_{11} =1$ in (a) confirmed by $y_1$ vs. $x_1$ plot in (b), AS between $y_2$ and $x_2$ for $\alpha_{22} =-1$ in (c) confirmed by $y_2$ vs. $x_2$ plot in $ (d)$  and, (e) chaotic driver $y_3$ and its response $x_3$ are in-phase state but the response is attenuated in (f) for $\alpha_{33} =0.5.$}
\end{figure}

We construct the  $H_1$ and $H_2$ matrices from the {\it Jacobian} matrices by following the rules as stated above: $H_1=[p_1 \;\;-1 \;\;0\;;\; 1\;\; a+p_2\;\; 0\;;\; 1\;\; 0\;\; p_3]$ and $H_2=[0 \;\;0 \;\;h_1\;; 0\;\; 0\;\; 0\;;\; 0\;\; 0\;\; h_2].$ By a suitable choice of the parameters $p_1=p_2=p_3=-3,$ $h_1=-1, h_2=0$ and $\gamma =1$, the eigenvalues of the $H$ matrix are determined as -9.6, -6.4, -5.4 which satisfies the Hurwitz criterion (8) and ensures stability of synchronization. This choice of parameters is certainly not unique, other choices of parameters are also available for which $H$ satisfies (8) as shown (dark region) in the $h_1 - p_1$ plane in Fig.3 where  $p_1=p_2=p_3$. By this equality, we find a symmetric non-synchronization region in white. However, other choices of parameters are always possible for $p_1 \not =p_2 \not=p_3$. Clearly, a wide choices of $h_1 - p_1$ parameters is available for a positive value of $\gamma$. Once the stability of synchronization is thereby established, we can easily target several synchronization scenarios, similar to the first example, by making different choices of the elements of the matrix $\alpha$ and coupling delay $\tau_c.$  If $\alpha_{11}=\alpha_{22}=\alpha_{33}=1.0,$ with no coupling delay ${(\tau_c=0)}$, systems (12) and (13) are in CS. It means all similar pairs of state variables of the driver and the response maintain a CS state. Similarly, for $\alpha_{11}= \alpha_{22}= \alpha_{33}=-1.0,$ all the pairs of state variables of the systems are in AS state.
\par {\it Mixed synchronization:} For numerical example, we consider a different choice of $\alpha_{11}=1, \alpha_{22}=-1, \alpha_{33}=0.5$ without a coupling delay to reveal an uncommon synchronization scenario. We find that different pairs of state variables of the driver and the response develop different types of coherence: $y_1(t)$ and $x_1(t)$ are in CS in Fig. 4(a), $y_2(t)$ and $x_2(t)$ in AS in Fig. 4(c). The $y(t)$ vs. $x(t)$ plots in Fig. 4(b) and 4(d) confirms CS, AS respectively in similar pairs of state variables. The third variable $y_3(t)$ of the driver in Fig. 4(e) is in-phase with the response variable $x_3(t)$ in Fig. 4(f) but attenuated by twice as seen in Fig. 4(e). This typical scenario is defined as a MS state where three different pairs of state variables of the systems (12) and (13) exist simultaneously in different coherent states. Note that the system (12) is not inversion symmetric, even then AS is realized in this system. The coupling thus overrules the restriction of inversion symmetry.

\begin{figure}
\includegraphics[width=80mm,height=100mm]{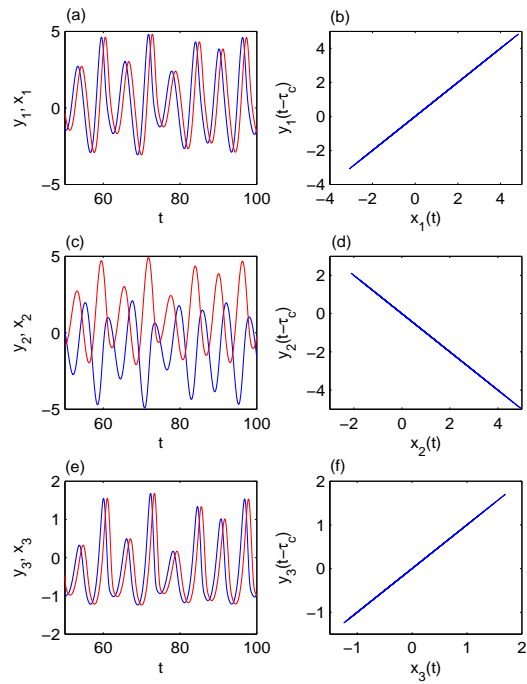}
\caption{\label . (Color online) Lag synchronization in delay R\"{o}ssler system: (a) LS for $\alpha_{11} =1,$ corresponding $x_1(t)$ vs. $y_1(t-\tau_c)$ plot in (b), (c) anti-LS for $\alpha_{22} =-1,$ corresponding $x_2(t)$ vs. $y_2(t-\tau_c)$ plot in (d0, (e) LS for $\alpha_{33} =1$, and $x_3(t)$ vs. $y_3(t-\tau_c)$ plot in (f).}
\end{figure}

\par In presence of a coupling delay, systems (12) and (13) will obviously develop a LS state. Instead, we present the MS scenario with LS for a coupling delay $\tau_c=1.0$. Let us choose $\alpha_{11}=1.0$, $\alpha_{22}=-1.0$ and $\alpha_{33}=1.0$ : three different pairs of state variables simultaneously develop LS, anti-LS and LS but with a common delay $\tau_c=1.0$. In Fig.5(a), the time series of $y_1(t)$ (online blue color) and $x_1(t)$ (online red color) are shown in LS. The corresponding $x_1(t)$ vs. $y_1(t- \tau_c)$ plot confirms LS in Fig.5(b). Figure 5(c) shows anti-LS in the pair of time series, $x_2(t)$  and $y_2(t-\tau_c)$ for $\alpha_{22}=-1.0$ and a coupling delay $\tau_c=1.0,$ which is confirmed by the $x_2(t)$ vs. $y_2(t-\tau_c)$ plot in Fig.5(d). In case of anti-LS, a pair of state variables are in opposite phase and amplitude correlated but shifted by the desired time lag $\tau_c$. The $x_3(t)$ and $y_3(t)$ in Fig.5(e) also shows LS since we choose $\alpha_{33}=1.0.$ Figure 5(f) plots $x_3(t)$ vs. $y_3(t-\tau_c)$ to confirm LS between $y_3(t)$ and $x_3(t)$. A scaling of the size of the driver attractor at response is also possible by making the elements of the $\alpha$ matrix larger/lower than unity. It is demonstrated that the choice of the elements of the $\alpha$ matrix is so flexible that one can easily target different desired synchronization scenarios and does not interfere in the stability of synchronization. By varying one or more elements of this $\alpha$ matrix, a practical scheme can be implemented for smooth control of synchronization from one regime to another without loss of synchronization during this transition. One can even externally force an abrupt change in one or more elements of the $\alpha$ matrix to induce a flip-flopping of synchronization which can be utilized for phase-shift keying [34] in digital communication. Next we use this property of the $\alpha$ matrix to demonstrate a GS scenario in the next section.

\begin{figure}
\begin{center}
\includegraphics[width=95mm,height=85mm]{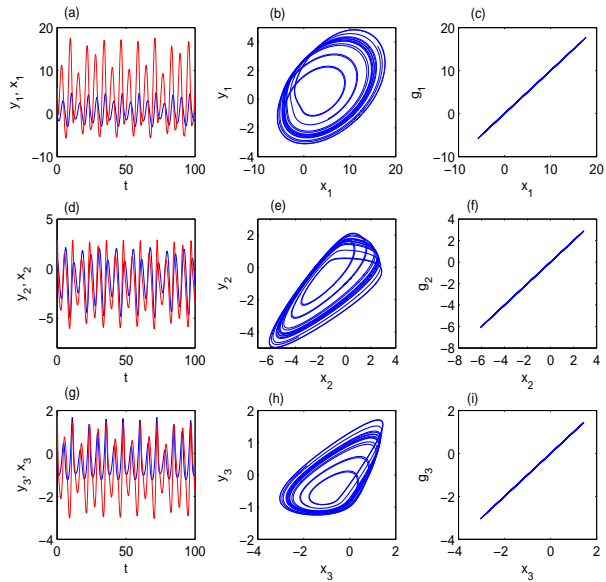}
\caption{\label. (Color online) Generalized synchronization in delay R\"{o}ssler system: (a) time series of driver $y_1$ (blue color) and response $x_1$ (red color), (b) shows $y_1$ vs $x_1$ and (c) transform driver $g_1$ and response variable $x_1$ in one-to-one relation. Similar plots of the pair ($y_2$, $x_2$), ($y_2$ vs.$x_2$), and  $g_2$ vs $x_2$ in (d)-(f). For the third response $g_3$, all similar plots are shown in (g)-(i)}.
\end{center}
\end{figure}

\section{Targeting generalized synchronization}
\par A GS scenario [28] was reported in nonidentical systems or largely mismatched systems under linear diffusive coupling. In a GS state the response system develop a functional relationship with the driver. GS in delay systems without coupling delay [35, 36] and with delay coupling [37] was also investigated, however, the coupling is assumed {\it a priori} known. Instead, we design the delay coupling for engineering a GS state [23] even in  identical or closely delay systems. We appropriately choose the elements of the $\alpha$ matrix for targeting a desired functional relationship between the driver and the response. As a simple example, we choose some of the off-diagonal elements of the $\alpha$-matrix as nonzero in addition to the principal diagonal elements, the driver-response system then enter into an interesting GS regime. For numerical demonstration, we use the coupled delayed R\"{o}ssler systems (12) and (13) with an arbitrary choice of the scaling matrix: $\alpha=[1\;\; -3\;\; -2\;;\;0\;\;1\;\;2\;;\;0\;\;0.5\;\;1]^T$. The goal dynamics $g(t)=[g_1(t)\;\; g_2(t)\;\; g_3(t)]^T$  is thereby defined by [$g_1(t)=y_1(t)-3y_2(t)-2y_3(t), g_2(t)=y_2(t)+2y_3(t), g_3(t)=0.5y_2(t)+y_3(t)]$. The state variables of the goal dynamics or the response variables are linear combinations of the driver variables since the elements of the $\alpha$ matrix are chosen constant or zero. Figure 6(a) shows time series of the driver $y_1(t)$ (online blue color) and the response $x_1(t)$ (online red color) which apparently shows no correlation by visual check. The $y_1(t)$ vs. $x_1(t)$ plot appeared to show no  correlation in Fig.6(b). However, the $x_1(t)$ vs. $g_1(t)$ plot in Fig.6(c) clearly shows 1:1 correlation and confirms that the targeted GS state is actually realized at the response system. Other pairs of state variables similarly show their corresponding targeted GS states defined by $g_2(t)$ and $g_3(t)$ in Fig.6(d)-(f) and Fig.6(g)-(i) respectively. Since the $\alpha$-matrix is only involved in the definition of a GS state it does not disturb the stability of synchronization. One can easily make a wider choices of the scaling matrix to define any desired functional relationship between the driver and response. A nonlinear functional relationship can also be adopted to target a more complex GS scenario by inserting state variables of the driver or a third dynamical system into the $\alpha$ matrix as shown, in detail, in case of instantaneous OPCL coupling for GS [23]. The proposed design scheme is thereby able to engineer a varieties of desired GS state in delay system too in presence of delay coupling. Note that different response variables develops a separate functional relation with corresponding driver variables which is basically a MS-GS regime. This is in contrast to the conventional GS scenario where all pairs of state variables of the coupled system achieve uniform functional relationship.

\section{Conclusion}
In conclusion, we introduced a design of delay coupling for synchronization of delay coupled systems. We established the stability conditions of synchronization by using the Krasovskii-Lyapunov function theory supported by the  Hurwitz matrix criterion. We achieved several important benefits: we can realize many synchronization regimes (CS, AS, LS, anti-LS, MS and GS) those are not usually realizable in all dynamical systems using the conventional coupling schemes. A scaling matrix is introduced in the coupling definition without disturbing the stability of synchronization. The scaling matrix allows enlarging or attenuating a driver attractor at a response system. A stabilization of the equilibrium origin of a response system as an AD or OD state is possible which has important application in practical delay system. Further, a MS and a MS-GS state can be engineered in slightly mismatched oscillators by a wider range of choices in the elements of the scaling matrix. The stability of synchronization is made independent of the system parameters that enables a flexible control of the synchronization. One can thus smoothly change any of the elements of the scaling matrix to induce a change from one synchronization regime to another without loss of stability during a transition. This is a special property [30] of this coupling design which can be used for practical applications.We supported the theory with numerical simulations of the Mackey-Glass system and a delayed R\"{o}ssler model.

\par S.K.D. acknowledges partial support by the DAE/BRNS, India under grant 2009/34/26/BRNS and I. G. is supported by the CNCSIS, Romania under Grant Nos. 12115/2008 and 2219/2009.


\end{document}